\documentclass[pra, twocolumn, showpacs, floatfix]{revtex4}  
\usepackage{hyperref}
\usepackage{amssymb}
\usepackage{amsmath}
\usepackage{float}
\usepackage{bbm}
\usepackage{graphicx}

\begin{document}

\title{Mott insulating phases and magnetism of fermions in a double-well optical lattice} 

\author{Xin Wang,$^{1}$ Qi Zhou,$^{1,2}$ and S. Das Sarma$^{1,2}$}

\affiliation{$^{1}$Condensed Matter Theory Center, Department of Physics,
University of Maryland, College Park, Maryland 20742, USA\\
$^{2}$Joint Quantum Institute, University of Maryland, College Park, Maryland 20742, USA}

\date{\today}

\begin{abstract}

We theoretically investigate, using non-perturbative strong correlation techniques, Mott insulating phases and magnetic ordering of two-component fermions in a two-dimensional double-well optical lattice. At filling of two fermions per site, there are two types of Mott insulators,  one of which is characterized by spin-$1$ antiferromagnetism below the Neel temperature. The super-exchange interaction in this system is induced by the interplay between the inter-band interaction and the spin degree of freedom.  A great advantage of the double-well optical lattice is that the magnetic quantum phase diagram and the Neel temperature can be easily controlled by tuning the orbital energy splitting of the two-level system. Particularly, the Neel temperature can be one order of magnitude larger than that in standard optical lattices, facilitating the experimental search for magnetic ordering in optical lattice systems.

\end{abstract}

\pacs{05.30.Fk, 67.85.-d, 71.27.+a, 03.75.-b}
\maketitle

\emph{Introduction}.- There are currently worldwide efforts in studying collective properties of cold atoms either in a single trap or in an optical lattice\cite{Bloch}. A central goal of these studies is to explore novel many-body quantum phases in both  bosonic and fermionic systems.  While both bosonic and fermionic Mott insulators have been realized in laboratories\cite{Bloch2, Chin, Esslinger, Bloch3}, the experimental search for magnetism in optical lattices is currently on-going. Most of these studies have been focusing on the single-band physics. For example,  it is known that two-component fermions in the lowest band can be used to study spin-$1/2$ antiferromagnetism\cite{Demler}. 

A question naturally arises: Is it possible to realize multi-band magnetic systems using cold atoms in optical lattices?  
Theoretical studies suggest, for example, exploring excited bands in optical lattices for searching novel magnetism\cite{Wu, Wu2}, partly because of the enhanced tunnel coupling in excited bands \cite{Zhai08,Sengstock11}.  
While there are currently experimental efforts along this direction to populate atoms in excited bands\cite{Bloch4, Hemmerich}, whether one can overcome the finite-life time problem of atoms in excited bands still remains unclear. 

On the other hand, there is a crucial practical issue on the energy and time scales of atoms in optical lattices. Ordinary optical lattices are characterized by extremely small energy scales,  which also lead to slow relaxation of lattice systems to equilibrium. For example, the tunneling of the lowest band is about a few nano-Kelvin which corresponds to a time scale of a few tens of milliseconds. The energy scale associated with the super-exchange interaction $t^2/U$ is even smaller since $U\gg t$ typically. As a result, the Neel temperature of antiferromagnetism in ordinary optical lattices is far too low for experimental observation. Meanwhile, it is also challenging for the system to reach equilibrium because of the long relaxation time. A scheme to enhance the relevant energy scales is therefore very desirable, particularly in the context of the experimental study of many-body magnetism in optical lattice systems.

In this Rapid Communication, we theoretically study quantum magnetism of fermions in a double-well optical lattice. Instead of the usual spin-$1/2$ magnetic ordering in an ordinary optical lattice,  the double-well effectively produces a spin-$1$ system.  The associated magnetism is induced by the inter-band interaction between the lowest two bands, and is a ground state property. Moreover, the characteristic energy scale for observing magnetism can be enhanced by one order of magnitude compared with the spin-$1/2$ magnetism in ordinary optical lattices.  As a result, the magnetism may be much easier to achieve and observe experimentally in the double-well optical lattice.

A double-well optical lattice contains two potential wells, which are separated by a barrier, on each lattice site. Its unique advantage is that the band gap between the lowest two bands is tunable\cite{Larson, Zhou}. When these two bands are very close to each other, interesting quantum many-body phenomena, which are completely absent in ordinary optical lattices, emerge\cite{Zhou, Zhou2}. We will see that the interplay between the orbital degree of freedom and the fermionic spin lies at the heart of the new physics reported here.  Theoretically it is however challenging to study fermions with spin degrees of freedom in the presence of multiple bands. We employ the dynamical mean-field theory (DMFT) method\cite{Georges96}, and study both the Mott insulating phases and magnetic properties of fermions in a double-well lattice. We show that at filling of two fermions per site,  the Mott insulator developed in the system can be either the triplet $(n_s,n_p)=(1,1)$ states  or the admixture $u(2,0)-v(0,2)$. For the former case, antiferromagnetic order emerges in the spin-1 channel, which should be experimentally observable.

\emph{Model}.- We consider the Hamiltonian containing a tight-binding-band part and an on-site interaction part, characterizing the two lowest bands (labeled by $s$ and $p$ respectively) in a symmetric double-well lattice, $H=H_{\rm band}+H_{\rm int}$.  In the real space, the band part can be written as
\begin{equation}
\begin{split}
H_{\rm band}=&\sum_{{\boldsymbol r}\sigma}\Big[(\varepsilon_s-\mu)s_{\sigma,{\boldsymbol r}}^\dagger s_{\sigma,{\boldsymbol r}}+(\varepsilon_p-\mu)p_{\sigma,{\boldsymbol r}}^\dagger p_{\sigma,{\boldsymbol r}}\Big]\\
&+\sum_{{\boldsymbol r}\sigma}\Big(-t_{sx}s_{\sigma,{\boldsymbol r}}^\dagger s_{\sigma,{\boldsymbol r}+{\boldsymbol x}}-t_{sy}s_{\sigma,{\boldsymbol r}}^\dagger s_{\sigma,{\boldsymbol r}+{\boldsymbol y}}\\
&+t_{px}p_{\sigma,{\boldsymbol r}}^\dagger p_{\sigma,{\boldsymbol r}+{\boldsymbol x}}-t_{py}p_{\sigma,{\boldsymbol r}}^\dagger p_{\sigma,{\boldsymbol r}+{\boldsymbol y}}+h.c.\Big),
\end{split}\label{band}
\end{equation}
where $s_{\sigma,{\boldsymbol r}}^\dagger$($p_{\sigma,{\boldsymbol r}}^\dagger$) creates a fermion with spin $\sigma$ on the $s$($p$) orbital of site ${\boldsymbol r}$,
$\varepsilon_s$ and $\varepsilon_p$ are the energies for $s$ and $p$ orbitals, and $\mu$ is the chemical potential. The hopping amplitude for $s$ and $p$ orbitals may differ in $x$ and $y$ directions, thus we label them by $t_{sx}$, $t_{sy}$, $t_{px}$, $t_{py}$ respectively.

\begin{figure}[t]
\begin{center}
\includegraphics[width=8.6cm, angle=0]{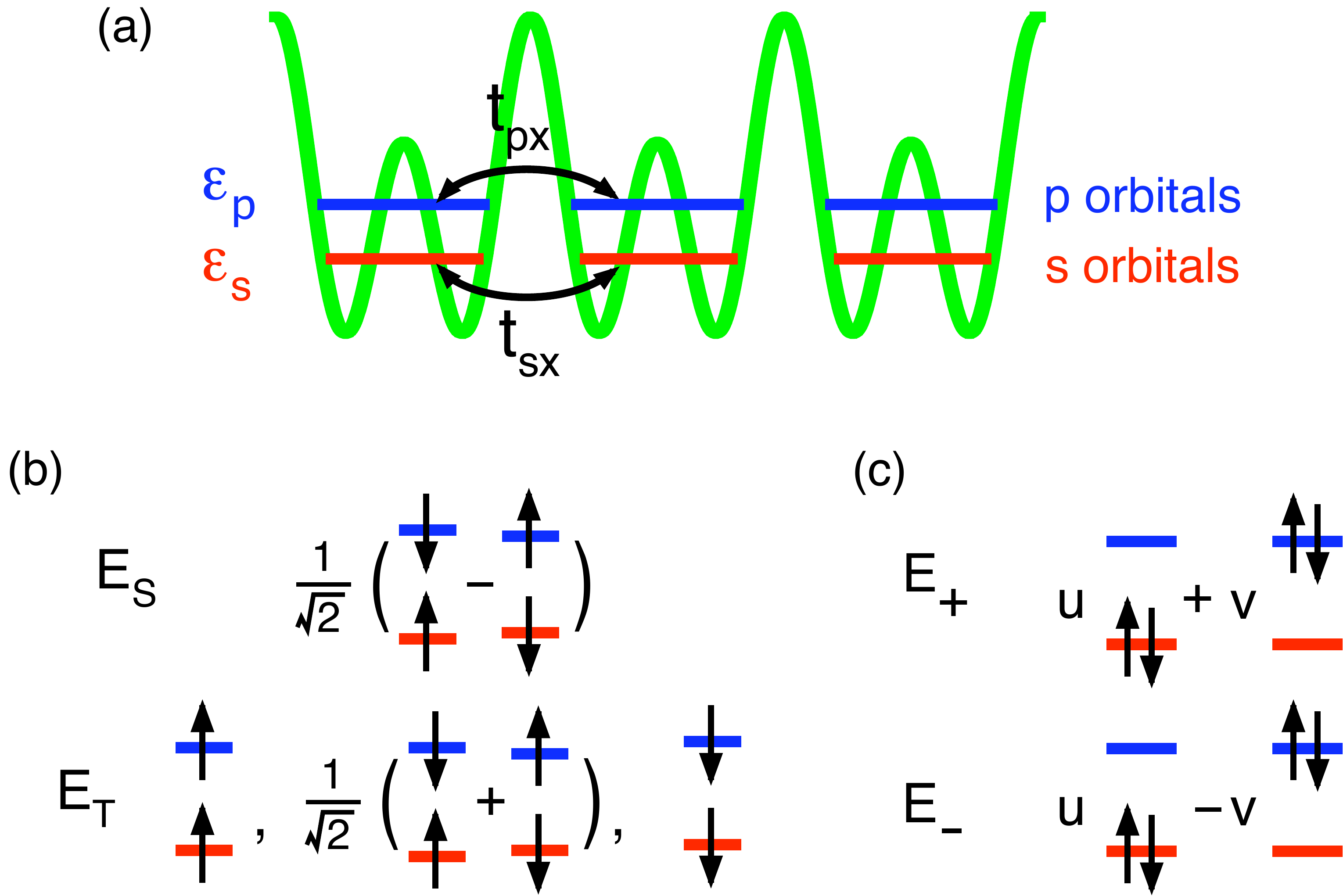}
\caption{(Color online) (a) Profile of the double-well potential (the green line) along the $x$-direction. The $s$ (red lines) and $p$ (blue lines) orbitals are schematically shown. The hopping integrals and energies for $s$($p$) orbitals, $t_{sx}$($t_{px}$) and $\varepsilon_s$($\varepsilon_p$), are indicated. Note that the potential is not double-welled along the $y$-direction (not shown), where the hopping integrals are $t_{sy}$ and $t_{py}$ correspondingly. (b) $(n_s,n_p)=(1,1)$ eigenstates in atomic limit, including triplet states with energy $E_T$ and a singlet state with energy $E_S$. (c) Linear combinations of $(2,0)$ and $(0,2)$ states as eigenstates for the Hamiltonian in atomic limit, whose energies are denoted by $E_\pm$. See the text for details.}\label{schematic}
\end{center}
\end{figure}

The interacting part of the Hamiltonian can be written as
\begin{equation}
\begin{split}
H_{\rm int}&=\sum_{\boldsymbol r}\Bigg[U_sn_{s\uparrow, {\boldsymbol r}}n_{s\downarrow, {\boldsymbol r}}+U_pn_{p\uparrow, {\boldsymbol r}}n_{p\downarrow, {\boldsymbol r}}\\
&+U_{sp}(n_{s\uparrow, {\boldsymbol r}}n_{p\downarrow, {\boldsymbol r}}+n_{s\downarrow, {\boldsymbol r}}n_{p\uparrow, {\boldsymbol r}})\\
&-U_{sp}\left(s_{\downarrow, {\boldsymbol r}}^\dagger p_{\uparrow, {\boldsymbol r}}^\dagger p_{\downarrow, {\boldsymbol r}}s_{\uparrow, {\boldsymbol r}}+p_{\uparrow, {\boldsymbol r}}^\dagger p_{\downarrow, {\boldsymbol r}}^\dagger s_{\uparrow, {\boldsymbol r}}s_{\downarrow, {\boldsymbol r}}+ h.c. \right)\Bigg],
\end{split}\label{int}
\end{equation} 
where $n_{\alpha\sigma,{\boldsymbol r}}=\alpha_{i\sigma,{\boldsymbol r}}^\dagger \alpha_{i\sigma,{\boldsymbol r}}$ ($\alpha=s,p$) is the number operator for orbital $\alpha$ at site ${\boldsymbol r}$, $U_\alpha=\frac{4\pi\hbar^2a_s}{M}\int d^3xW^4_\alpha({\bf x})$ denotes the intra-band interaction, while $U_{sp}=\frac{4\pi\hbar^2a_s}{M}\int d^3xW_s^2({\bf x})W_p^2({\bf x})$ denotes the inter-band interaction, where $a_s$ is the scattering length, $M$ is the mass of the fermion, and $W_\alpha({\bf x})$ is the Wannier wave function for each band. The inter-orbital terms in Eq.(\ref{int}) characterized by $U_{sp}$ are referred as density-density, spin-exchange and pair-hopping interaction. This model is essentially the rotationally invariant Slater-Kanamori interaction widely studied in 
transitional metal oxides \cite{Wang11}. The main difference here
is that the spin-exchange and pair-hopping are as strong as the inter-orbital density-density interaction. An important parameter which controls the multi-band physics is the energy level splitting between the two levels, defined as $\Delta\equiv\varepsilon_p-\varepsilon_s$. When $\Delta$ is small or intermediate, interactions between the two orbitals give rise to interesting phenomena, as we discuss below. When $\Delta$ becomes very large the physics reduces to that of the single-band model. The Hamiltonian in Eq.~\eqref{int} has been previously considered for fermions at resonance in an ordinary optical lattice in one-dimension\cite{AHo}. In our case, the reduced band gap makes the realization of a two-band system more practical in current experiments. Moreover, higher dimensionality of our system gives distinct physical phenomena not accessible in one dimension.

We start from the atomic limit, where the tunneling terms are absent. We are interested in the states at filling of two fermions per site, the schematics of which are shown in Fig.~\ref{schematic}(b) and 1(c). When there is one fermion in each orbital, they form triplets which is denoted as $(n_s,n_p)=(1,1)$, namely, $p_{\uparrow}^{\dagger}s_{\uparrow}^{\dagger}\left|0\right\rangle$,
$\frac{1}{\sqrt{2}}\left(p_{\uparrow}^{\dagger}s_{\downarrow}^{\dagger}+p_{\downarrow}^{\dagger}s_{\uparrow}^{\dagger}\right)\left|0\right\rangle$, and
$p_{\downarrow}^{\dagger}s_{\downarrow}^{\dagger}\left|0\right\rangle$, with degenerate energy $E_T=2(\varepsilon_s-\mu)+\Delta$. The singlet state
$\frac{1}{\sqrt{2}}\left(p_{\uparrow}^{\dagger}s_{\downarrow}^{\dagger}-p_{\downarrow}^{\dagger}s_{\uparrow}^{\dagger}\right)\left|0\right\rangle$ has a higher energy $E_S=2(\varepsilon_s-\mu)+\Delta+2U_{sp}$.  $E_S>E_T$ simply because two fermions interact  with each other by $s$-wave short-range interaction. In the spirit of the Hubbard model, atoms with different spins repel with each other and atoms with the same spin do not interact. On the other hand, the two fermions can also form admixtures $u(2,0)\pm v(0,2)$, as shown in Fig.~\ref{schematic}(c), due to the pair-hopping interaction.  The eigenenergies are $E_{\pm}=2(\varepsilon_s-\mu)+\Delta+\frac{U_p+U_s}{2}\pm\sqrt{\left(\Delta+\frac{U_p-U_s}{2}\right)^2+U_{sp}^2}$. By controlling $\Delta$ in the double-well lattice, $E_{-}$ can be made either smaller or larger than $E_T$.  
Throughout the paper we fix parameters $U_s=12t$, $U_p=14t$, $U_{sp}=12t$, and vary $\Delta$ and the temperature $T$. Straightforward algebra reveals that at the critical value $\Delta_c=4t$, $E_{-}=E_T$. For latter use we note that the hopping integrals are chosen as $t_{sx}=t_{sy}=t_{py}=t$, $t_{px}=2t$. The large $t_{px}$ stems from the fact that $p$ bands are spatially more extended along the $x$-direction. However, our solution to the lattice model as well as the physics therein does not depend on this particular set of parameters in any important way.

When the hopping terms are switched on, we employ the single-site DMFT \cite{Georges96} to solve the strongly-correlated interacting lattice fermion problem. The key approximation is the neglecting of momentum dependence of the self-energy: ${\bf \Sigma}({\boldsymbol k}, \omega)\rightarrow{\bf \Sigma}(\omega)$, which is solved iteratively from an auxiliary quantum impurity problem plus a self-consistency condition. We use the matrix representation of the continuous-time hybridization-expansion quantum Monte Carlo impurity solver \cite{Werner06} prescribed specifically for multi-band interactions. This is a state-of-the-art highly demanding numerical solution of the strongly interacting multi-band lattice Hubbard model in the context of our double-well optical lattice system.

\begin{figure}[]
\begin{center}
\includegraphics[width=8.4cm, angle=0]{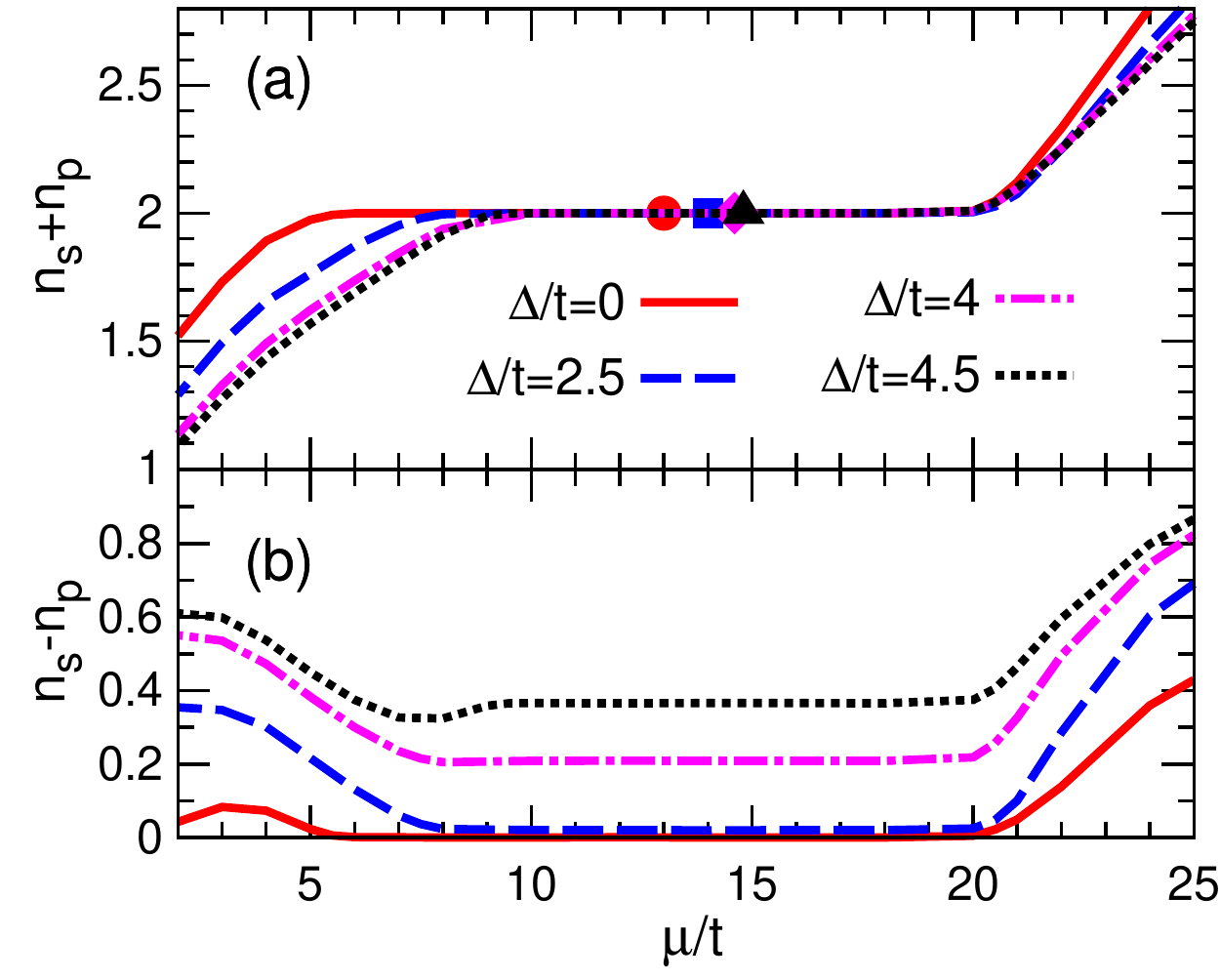}
\caption{(Color online) (a) Total occupancy ($n_s+n_p$) versus chemical potential $\mu$, calculated for three different values of $\Delta$. The calculation is done without magnetic order at temperature $T=0.2t$. $\varepsilon_s=0$ in this plot. The points (the circle, the square the diamond and the triangle) indicate the location at approximately the center of the gap for the line with corresponding color, where we study magnetic ordering. (b) The difference in occupancy $n_s-n_p$ plotted at the same $\mu$ scale.}\label{nvsmu}
\end{center}
\end{figure}

\emph{Mott physics}.- There are multiple choices to fill a single lattice site with two fermions, forming different types of Mott insulator. To distinguish them, we have calculated both $n_s+n_p$ and $n_s-n_p$ as functions of $\mu$ for different values of $\Delta$, as shown in Fig.~\ref{nvsmu}(a). A Mott insulating gap at filling two is evident for all cases, and there is no qualitative difference in the value of  $n_s+n_p$ between difference cases.  However, the difference in occupancy for the two orbitals ($n_s-n_p$) shows distinct behaviors. At very large level splitting $\Delta=4.5t$, $n_s\gg n_p$. This is consistent with the analysis in the previous section for the atomic limit, where each lattice site is filled by the state $u(2,0)-v(0,2)$ and $u\gg v$.  In contrast, at small $\Delta$, e.g. as seen in Fig.~\ref{nvsmu}(b) for $\Delta=2.5t$ and $\Delta=0$, $n_s\approx n_p$ in the Mott insulating regime. This indicates that on each site the triplet states dominate the ground state. This is also consistent with the atomic limit where the energy of (1,1) states $E_T$ continuously decreases and eventually the (1,1) triplet becomes the ground state with decreasing $\Delta$, as discussed in the previous section. The transition between the two types of insulator is a crossover. Since in this paper we focus on properties at non-zero temperature, we shall not discuss the nature of this transition at zero temperature.

For $\Delta\rightarrow\infty$, the magnetism is manifestly absent and the ground state continuously connects to the trivial band insulator in the lowest band of an ordinary optical lattice.  For small and intermediate $\Delta$, however, a magnetization of spin-$1$ may arise from the triplet states on a single lattice site.  As a result, the physics of magnetic ordering in double-well lattices at filling two is far richer than that in standard optical lattices.

\emph{Magnetic order}.-  When the (1,1) states dominate the on-site Fock states for small $\Delta$, the interacting Hamiltonian can be mapped to a spin-1 Heisenberg model, which can be written as
\begin{equation}
H_{\rm eff}=\sum_{\boldsymbol r}\left(J_x{\boldsymbol S}_{\boldsymbol r}\cdot{\boldsymbol S}_{{\boldsymbol r}+{\boldsymbol x}}+J_y{\boldsymbol S}_{\boldsymbol r}\cdot{\boldsymbol S}_{{\boldsymbol r}+{\boldsymbol y}}\right),\label{heisen}
\end{equation}
where
\begin{equation}
J_x=\frac{2t_{sx}^2}{U_s+U_{sp}}+\frac{2t_{px}^2}{U_p+U_{sp}},\ J_y=\frac{2t_{sy}^2}{U_s+U_{sp}}+\frac{2t_{py}^2}{U_p+U_{sp}},\label{Jvalue}
\end{equation}
${\boldsymbol S}_{\boldsymbol r}=A^\dagger {\boldsymbol \Sigma} A$ is a spin-1 operator, ${\boldsymbol \Sigma }$ is spin-1 Pauli matrices, and $A=(\Psi^\dagger_1, \Psi^\dagger_0, \Psi^\dagger_{-1} )^T$ are creation operators for triplet states $p_{\uparrow}^{\dagger}s_{\uparrow}^{\dagger}\left|0\right\rangle$,
$\frac{1}{\sqrt{2}}\left(p_{\uparrow}^{\dagger}s_{\downarrow}^{\dagger}+p_{\downarrow}^{\dagger}s_{\uparrow}^{\dagger}\right)\left|0\right\rangle$, and $p_{\downarrow}^{\dagger}s_{\downarrow}^{\dagger}\left|0\right\rangle$. Physically, the spin-exchange terms in Eq.~\eqref{heisen} come from the exchange of fermions with different spins between the nearest-neighbor sites in either of the two orbitals. Both orbitals contribute to the spin-exchange terms in the effective Hamiltonian. 

In the one-dimensional case, Eq.~\eqref{heisen} has previously been derived in Ref.~\onlinecite{AHo}. For that case, it has been known that the one-dimensional spin-1 chain does not have any magnetic order, rather the Haldane phase. Nevertheless, a two-dimensional spin-1 system can develop antiferromagnetic ground states\cite{2Dspin1,2Dspin2,2Dspin3}. Therefore, one expects to see antiferromagnetic spin ordering in a double-well optical lattice when the gap $\Delta$ is small and the temperature is low\cite{Mermin}.

\begin{figure}[]
\begin{center}
\includegraphics[width=8.2cm, angle=0]{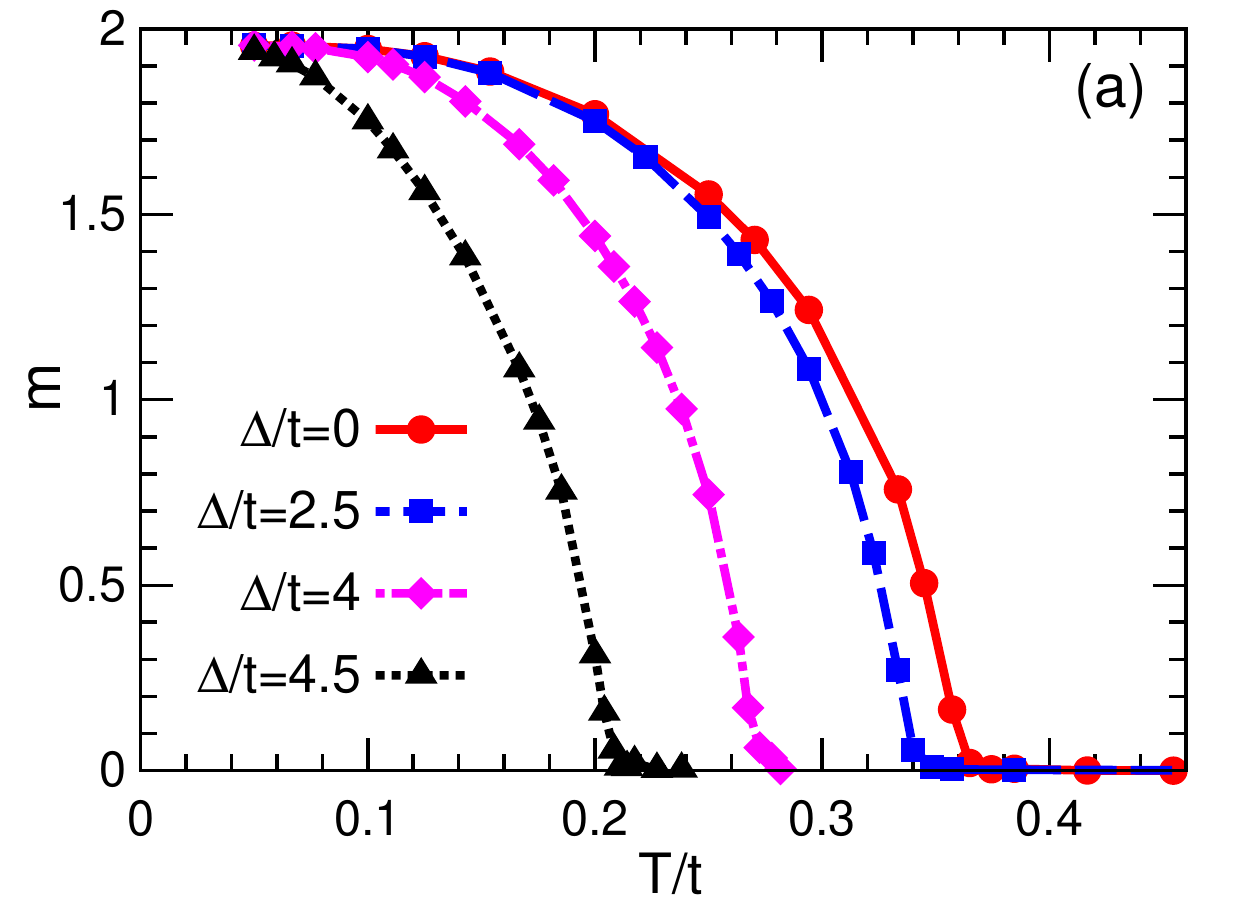}
\includegraphics[width=8.4cm, angle=0]{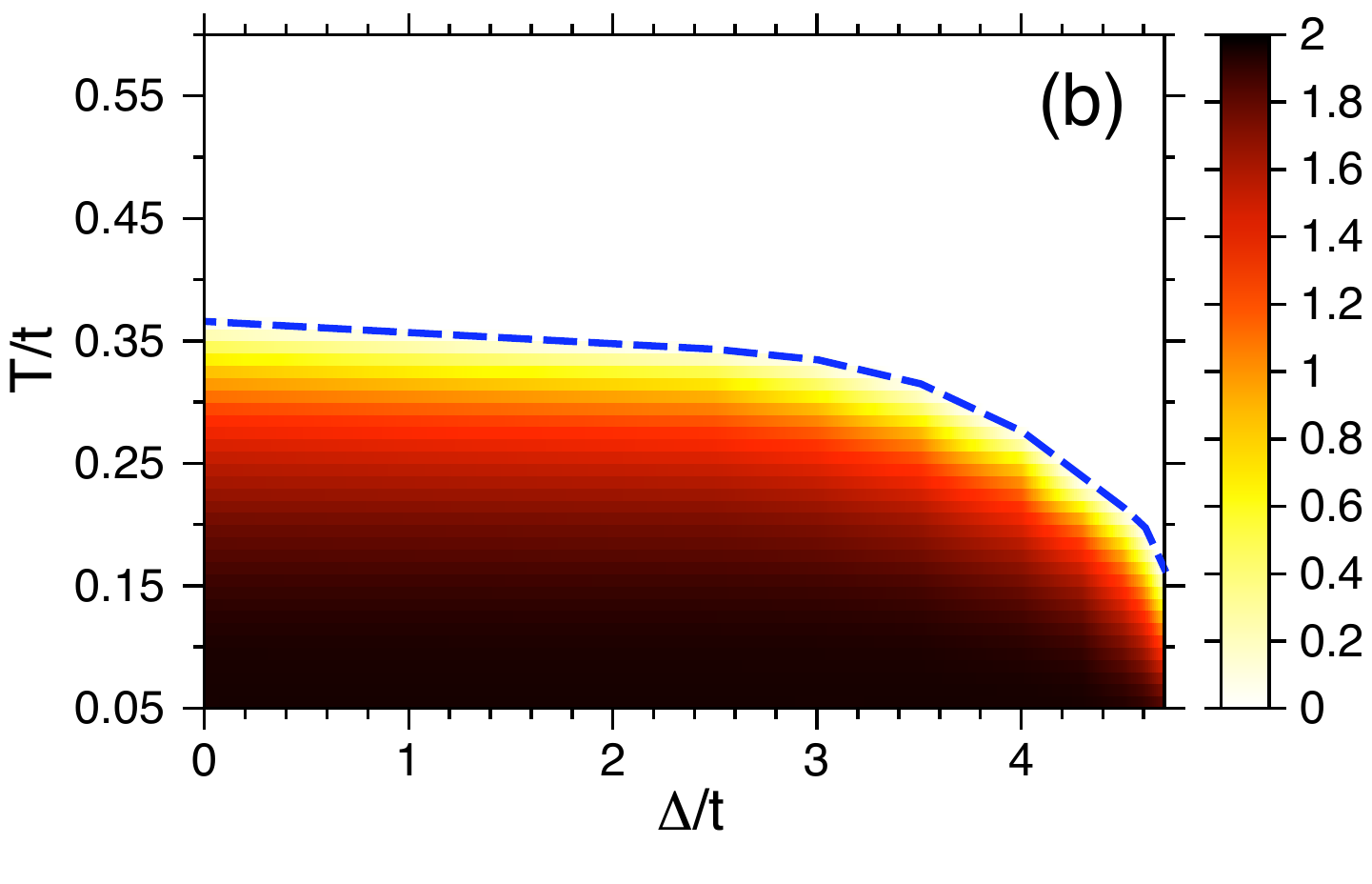}
\caption{(Color online) (a) Total magnetization $m=\left|n_{s\uparrow}-n_{s\downarrow}+n_{p\uparrow}-n_{p\downarrow}\right|$ versus temperature, for four selective values of $\Delta$. The chemical potential is selected at approximate center of the gap (see Fig.~\ref{nvsmu}). (b) Color plot of the magnetization on the $\Delta$-$T$ plane. In the white regime there is no magnetic order, while for low $T$ the magnetization reaches the maximum value, indicated as dark colors. The Neel temperature is shown as blue dashed lines seperating colored and white regimes. Note that the $y$-axis does not start from zero: it starts from the lowest temperature $T=0.05t$ reached in the DMFT calculation.}\label{mvsT}
\end{center}
\end{figure}

To characterize the magnetization, we define $m=\left|n_{s\uparrow}-n_{s\downarrow}+n_{p\uparrow}-n_{p\downarrow}\right|$ and solve the full Hamiltonian $H=H_{\rm band}+H_{\rm int}$.  The results for $m$ as a function of the temperature for different $\Delta$ are shown in Fig.~\ref{mvsT}(a). Clearly, the magnetization arises below the Neel temperature (denoted by $T^{\rm Neel}$) and saturates to its maximum value as the temperature approaches zero. For $\Delta=0$, the antiferromagnetic ordering is most pronounced: it has the highest Neel temperature $T^{\rm Neel}\simeq 0.37t$. However, this must be interpreted with caution because experimentally it is very difficult to tune the two bands overlap with each other while keeping the tight-binding model valid. We therefore focus on the cases with nonzero $\Delta$. As $\Delta$ is increased, the magnetization drops faster as $T$ increases, and the Neel temperature decreases. For a relatively large $\Delta=4.5t$, $T^{\rm Neel}\simeq 0.21t$.  Note that this value of $\Delta$ is already above the critical value $\Delta_c=4t$ in the atomic limit where $u(2,0)-v(0,2)$ is the ground state. This indicates that many-body effects, such as the correlation between nearest-neighbor sites, enhance the threshold of $\Delta_c$ for a finite $m$ to emerge.

To give a broader picture, we show in Fig.~\ref{mvsT}(b) a color plot of the magnetization on a plane, of which the axes are the energy level splitting $\Delta$ and the temperature $T$. The blue dashed line, separating colored and white regimes, indicates the Neel temperature, above which no magnetic order is present. The dark color shown near the lowest accessible temperature $T=0.05t$ in our simulations characterizes the saturation of the magnetization to its maximum value. As the temperature is increased to intermediate values, the color turns to red, indicating a moderate drop of the magnetization. When the temperature is close to the Neel temperature, the magnetization drops rapidly, as can be seen from the narrow yellow edge. A close examination of the Neel temperature reveals that it drops relatively slowly for $\Delta<\Delta_c$, but very rapidly for $\Delta>\Delta_c$. This is consistent with the qualitative atomic picture. 
For $\Delta\rightarrow\infty$, the magnetic ordering and the corresponding Neel temperature would eventually vanish. However, we have shown that the Neel ordering would survive at reasonably large values of $\Delta$. This is remarkable, because previous analytical argument of mapping to the spin-1 model\cite{AHo} is valid for $\Delta\rightarrow0$ only, while for non-zero $\Delta$ a direct numerical solution to the lattice model Eqs.~\eqref{band} and \eqref{int} is highly nontrivial. Our results relieve the restriction posed on experiments where achieving very small $\Delta$ is difficult.
 An alternative way to appreciate these results is that both $m$ and $T^{\rm Neel}$ are tunable by controlling $\Delta$, which has obvious important experimental implications.

\emph{Enhancement of super-exchange interaction}.- The increase of $J$ [cf. Eqs.~\eqref{heisen} and \eqref{Jvalue}] in a double-well lattice comes from two sources.  First, as seen from Eq.~\eqref{Jvalue}, in addition to $t_{s}$, $t_{px}$ and $t_{py}$ also enter the expression for the super-exchange interaction $J$. The large value of $t_{px}$ then enhances the amplitude of
$J$, similar to the antiferromagnetism arising from $p$ bands alone \cite{Zhai08}. Second and most importantly, $t_{sx}$ itself is significantly enhanced in a double-well optical lattice. It has been shown that $t_{sx}$ can be increased by one order of magnitude at a given lattice depth for some realistic experimental parameters\cite{Zhou}.  Thanks to the potential barrier in the center of each lattice site of a double-well lattice, the Wannier wave function of the lowest band spreads its weight toward the edge of the corresponding unit cell, which consequently
enhances the overlap between Wannier wave functions on adjacent sites,
leading to an increase in the tunneling amplitude.
As a result, the Neel temperature can be strongly enhanced, easily by one order of magnitude. The larger energy scale associated with the tunneling and super-exchange interaction will also help to reach equilibrim faster in the strongly interacting region. We emphasize that this spin-1 antiferromagntism originates from the unique feature of the double-well optical lattice: The $s$ and $p$ bands can be tuned close to each other, and the resulting magnetic ordering incorporates both bands. It is this feature that distinguishes our theory from previous proposals regarding $p$ bands alone\cite{Wu2, Zhai08}.

\emph{Conclusion}.- Using non-perturbative `DMFT with continuous-time quantum impurity solver' direct numerical techniques, we study two-component fermions in a double-well square optical lattice, with two interacting orbitals per site. The Mott insulator at filling two is constituted either by triplet $(n_s, n_p)=(1,1)$ or an admixture $u(2,0)-v(0,2)$. For the one associated with the triplets, antiferromagnetic order emerges in the spin-1 channel below the Neel temperature, which is determined by the energy splitting between the two orbitals and the tunneling amplitude. We establish that, as $t_{p}$ contributes to $J$ and $t_{sx}$ is significantly enlarged in double-well lattices, the Neel temperature can be one order of magnitude larger than that of the one-band system in ordinary optical lattices, thus perhaps enabling the direct experimental observation of the elusive Neel antiferromagnetism in cold atomic systems. Our work should facilitate the search of magnetic order in optical lattice systems.

\emph{Acknowledgements:} We thank M. Cheng and A. J. Millis for discussions. This work is supported by JQI-NSF-PFC, ARO-DARPA-OLE, JQI-ARO-MURI, and JQI-AFOSR-MURI.  The impurity solver in the DMFT procedure is based on a code primarily developed by P. Werner, and uses the ALPS library\cite{ALPS}.

\end{document}